\documentclass[aps,pra,preprint,amsmath]{revtex4-1}
\usepackage{graphicx,dcolumn,bm,color,microtype,hyperref}

\newcommand{\br}{\mathbf{r}}
\newcommand{\bu}{\mathbf{u}}
\newcommand{\bo}{\mathbf{0}}
\newcommand{\Ec}{\varepsilon_c}
\newcommand{\eps}{\varepsilon}
\newcommand{\EHF}{\varepsilon_\text{HF}}

\newcommand{\mc}{\multicolumn}

\newcommand{\mEh}{\text{m}E_\text{h}}

\newcommand{\rs}{r_s}

\newcommand{\db}[2]{\langle #1 || #2 \rangle}
\newcommand{\alert}[1]{\textcolor{black}{#1}}

\newcolumntype{d}[1]{D{.}{.}{#1}}

\DeclareMathOperator{\Ci}{Ci}

\begin{document}

\title{Generalized local-density approximation and one-dimensional finite uniform electron gases}	

\author{Pierre-Fran{\c c}ois Loos}
\email{Corresponding author: pf.loos@anu.edu.au}
\affiliation{Research School of Chemistry, Australian National University, Canberra ACT 0200, Australia}

\begin{abstract}
We explicitly build a generalized local-density approximation (GLDA) correlation functional based on one-dimensional (1D) uniform electron gases (UEGs). 
The fundamental parameters of the GLDA \textemdash~a generalization of the widely-known local-density approximation (LDA) used in density-functional theory (DFT) \textemdash~are the electronic density $\rho$ and a newly-defined two-electron local parameter called the hole curvature $\eta$. 
The UEGs considered in this study are finite versions of the conventional infinite homogeneous electron gas and consist of $n$ electrons on an infinitely thin wire with periodic boundary conditions. We perform a comprehensive study of these finite UEGs at high, intermediate and low densities using perturbation theory and quantum Monte Carlo calculations.  
We show that the present GLDA functional yields accurate estimates of the correlation energy for both weakly and strongly correlated one-dimensional systems and can be easily generalized to higher-dimensional systems.
\end{abstract}


\maketitle
%---------------------------------------------
\section{Introduction}
%---------------------------------------------

%---------------------------------------------
\subsection{Local-density approximation}
%---------------------------------------------
The local-density approximation (LDA) is the simplest approximation within density-functional theory (DFT) \cite{ParrBook}. 
It was first introduced by Kohn and Sham (KS) in 1965 to treat systems with slowly varying density \cite{Kohn65}. 
It assumes that a real, non-uniform system can be treated as a collection of infinitesimally-small uniform electron gases (UEGs) of density $\rho$. 
Thus, in principle, if one knows the reduced (i.e. per electron) correlation energy $\Ec$ of the UEG for any density $\rho$, 
one is able, by suming the individual contributions over all space, to obtain the LDA correlation energy
\begin{equation}	
\label{EcLDA}
	E_\text{c}^\text{LDA} = \int \rho(\br) \,\Ec[\rho(\br)] \,d\br.
\end{equation}
In the LDA, the correlation functional $\Ec[\rho(\br)]$ is based on the infinite UEG (IUEG) paradigm \cite{Fermi26, Thomas27}  \textemdash~a model built by allowing the number $n$ of electrons in a volume $V$ to approach infinity with $\rho = n/V$ held constant \textemdash~and analytically constructed \cite{VWN80, Perdew81, Perdew92, Sun10} by combining accurate quantum Monte Carlo (QMC) calculations \cite{Ceperley78, Ceperley80} and exact results from the high-density \cite{Wigner34, Macke50, Bohm53, Pines53, GellMann57, DuBois59, Carr64, Misawa65, Onsager66, Isihara75, Isihara76, Wang91, Hoffman92, Endo99, Ziesche05, 3DEG11} and low-density \cite{Maradudin60, Carr61, Fein61} limits.

Although it describes molecular bonding reasonably well compared to the Thomas-Fermi model \cite{Fermi26, Thomas27, Gunnarsson77}, this rather crude approximation had mixed success \cite{ParrBook}. 
Its main drawback is due to the large error in correlation energies, which are too large roughly by a factor two \cite{Tong66}. 
However, a subtle compensation of error between exchange and correlation makes the total energy usually in good agreement with experimental results \cite{Ernzerhof97}. 

%----------------------------------------------------------------
\subsection{Generalized local-density approximation}
%----------------------------------------------------------------
The birth of the generalized local-density approximation (GLDA) supervenes from the observation that a UEG of density $\rho$ is not uniquely defined \cite{UEGs12}. 
Indeed, we showed that two UEGs with same density can have different correlation energies depending on the UEG ``finiteness''. 
Thus, inspired by a number of previous researchers \cite{Colle75, Stoll80, Becke83, Luken84, Dobson91}, we introduced a new local two-electron parameter $\eta$ measuring the tightness of the correlation hole around an electron \cite{GLDA1}. 

From a practical point of view, at the Hartree-Fock (HF) or KS level of theory, the HF/KS dimensionless hole curvature is, in $D$ dimensions, 
\begin{equation} 
\label{eta-def}
	\eta(\br) = \frac{\tau(\br) - \tau_\text{W}(\br)}{\tau_\text{IUEG}(\br)} \propto \frac{\nabla^2_\bu P(\bo|\br)}{\rho(\br)^{2/D+1}} ,
\end{equation}
where the HF/KS conditional intracule 
\begin{equation}
	P(\bu|\br) = \frac{\rho_2(\br,\br+\mathbf{u})}{\rho(\br)}
\end{equation} 
measures the proximity of other electrons to one at $\br$, and $\rho_2(\br_1,\br_2)$ is the diagonal part of the second-order density matrix \cite{DavidsonBook}. 
In Eq.~\eqref{eta-def}, 
\begin{equation}
	\tau(\br) = \frac{1}{2} \sum_a^\text{occ} | \nabla\psi_a |^2
\end{equation}
is the HF/KS kinetic energy density, 
\begin{equation}
	\tau_\text{W}(\br) = \frac{|\nabla\rho(\br)|^2}{8\,\rho(\br)}
\end{equation} 
is the von Weizs{\"a}cker kinetic energy density, and 
\begin{equation}
	\tau_\text{IUEG}(\br) = \frac{\pi  D}{D/2+1}\Gamma \left(D/2+1\right)^{2/D} \rho(\br)^{2/D+1} 
\end{equation}
is the kinetic energy density of the (ferromagnetic) IUEG \cite{Glomium11} (where $\Gamma(x)$ is the gamma function \cite{NISTbook}).
Because the $\eta$ parameter defined in \eqref{eta-def} requires only the orbitals and their gradient, it can be easily evaluated with conventional quantum chemistry softwares.
We note that $\eta$ is also related to the inhomogeneity parameter used to construct exchange functionals within the meta-generalized gradient approximation \cite{Tao03, Perdew09, Sun12}.
Using the ``exact'' hole curvature would be probably more appropriate. However, this quantity is hardly accessible except for small systems \cite{GLDA1}.

Although the choice of the one- and two-electron parameters is not uniquely defined, in combination with the usual one-electron density parameter $\rho$, the hole curvature $\eta$ provides an unambiguous definition of a UEG, and the GLDA correlation energy is \cite{GLDA1}
\begin{equation}
\label{EcGLDA}
	E_\text{c}^\text{GLDA} = \int \rho(\br) \,\Ec[\rho(\br),\eta(\br)] \,d\br.
\end{equation}
Like the LDA, the GLDA can be used for atomic, molecular and periodic systems.

Our first investigation of the GLDA performance for one-dimensional (1D) systems \cite{GLDA1} showed very promising results using a non-self-consistent post-HF procedure \footnote{By ``non-self-consistent post-HF calculation'', we mean that we are using the converged HF density (i.e. obtained at the end of the HF calculation) to calculate the LDA and GLDA correlation energy based on \eqref{EcLDA} and \eqref{EcGLDA}.}.
In Ref \cite{GLDA1}, we built a GLDA correlation functional $\Ec[\rho(\br),\eta(\br)]$ based on the ``electrons-on-a-ring'' paradigm (also called ringium) which corresponds to electrons on a ring interacting \textit{through} the ring \cite{QuasiExact09, QR12, Ringium13} (see Appendix \ref{app:ringium}).

In the present study, we propose to follow a more conventional approach and explicitly build a GLDA correlation functional based on electrons on an infinitely thin wire with periodic boundary conditions (PBC) \cite{Astrakharchik11, Lee11a, UEG1D}. 
These UEGs can be seen as finite versions of the usual IUEG, and feel particularly attractive in the development of GLDA functionals. 
Note that the 1D IUEG at intermediate densities has been intensively studied by Lee and Drummond \cite{Lee11a}, while its high-density correlation energy has been studied in details in Ref.~\cite{UEG1D}. 
One of the aim of the present manuscript is to compare ring-based and wire-based models to build 1D GLDA functionals. 
Because UEGs can be created using various external potentials, it is important to know which of these paradigms is the more appropriate to model ``realistic'' systems.
A comparison between the electrons-on-a-wire and electrons-on-a-ring models is given in Appendix \ref{app:ringium}.

PBC are usually an issue in calculations on extended quantum systems because they inevitably introduce ``finite-size errors'', one of the major limitations of the application of accurate many-body techniques to periodic systems \cite{Fraser96, Kent99, Lin01, Kwee08, Drummond08, Ma11}. 
However, in this Article, we propose to take advantage of these finite-size errors to build finite UEGs. 

In order to gather information to build the GLDA functional in Sec.~\ref{sec:GLDA}, we perform a comprehensive study of these finite UEGs for high, intermediate, and low densities using perturbation theory and quantum Monte Carlo calculations (Sec.~\ref{sec:UEG}).  
In Sec.~\ref{sec:res}, we show that this new GLDA correlation functional yields accurate correlation energies for both weakly and strongly correlated 1D systems. 
We discuss its generalization to higher-dimensional systems in Sec.~\ref{sec:ccl}. 
Atomic units are used throughout and correlation energies are reported in millihartree ($\mEh$).

%----------------------------------
\section{\label{sec:UEG}
Finite uniform electron gases}
%----------------------------------
We consider UEGs composed of $n$ electrons with coordinates $x_i$ on an infinitely thin wire of length $L$ with PBC \cite{Lee11a}. 
For sake of mathematical simplicity, we map the spatial dimension onto a ring of radius $R = L/2\pi$ with coordinates $\theta_i = 2\pi\,x_i/L  \in [-\pi,+\pi]$ and interelectronic distances $\theta_{ij} = \left| \theta_i - \theta_j \right| \in [0,\pi]$. 
Thus, the PBC are naturally fulfilled, and the uniform electronic density $\rho$ (or equivalently the Wigner-Seitz radius $r_s$) is 
\begin{equation}
	\rho = \frac{n}{L} = \frac{n}{2\pi R} = \frac{1}{2\,r_s}.
\end{equation}
The Hamiltonian of a $n$-electron UEG is $\Hat{H} = \Hat{T} +\Hat{V}$, where the kinetic and potential operators are
\begin{align}
\label{H-wire}
	\Hat{T} & = - \frac{1}{2} \sum_{i=1}^{n} \frac{\partial^2}{\partial x_i^2} =  - \frac{1}{2R^2} \sum_{i=1}^n \frac{\partial^2}{\partial \theta_i^2},
	&
	\Hat{V} & = \sum_{i<j}^{n} v(x_{ij}) = \frac{1}{R}\sum_{i<j}^{n} v(\theta_{ij}).
\end{align}
The non-interacting orbitals and their corresponding energies are
\begin{align}
\label{orbitals}
	\psi_m(\theta) & = \frac{e^{i\,m\,\theta}}{\sqrt{2\pi R}},
	&
	\kappa_m & = \frac{m^2}{2R^2},
\end{align}
with 
\begin{equation}
	m = \begin{cases}
			\ldots, -2, -1, 0, +1, +2, \ldots,							&	\text{if $n$ is odd,}	\\
			\ldots, -\frac{3}{2}, -\frac{1}{2}, +\frac{1}{2}, +\frac{3}{2}, \ldots,	&	\text{if $n$ is even.}
		\end{cases}
\end{equation}
while the first-order density matrix is given by \cite{DavidsonBook}
\begin{equation}
	\rho_1(\theta_{12}) = \sum_a^\text{occ} \psi_a^*(\theta_1) \psi_a(\theta_2) = \frac{n}{\rho} \frac{\sin\left( n\,\theta_{12}/2\right)}{{\sin\left( \theta_{12}/2\right)}},
\end{equation}
where the summation over the occupied orbitals is 
\begin{equation}
	a = - \frac{n-1}{2}, - \frac{n-3}{2}, \ldots,  +\frac{n-3}{2}, +\frac{n-1}{2}.
\end{equation}

Due to the PBC, the electron $i$ interacts with electron $j$ and all of electron $j$'s periodic images 
\footnote{The self-interaction of electron $i$ with its own periodic images is part of the Madelung energy, which is constant. Thus, it can be omitted.}, 
and the Ewald interaction potential is \cite{Saunders94, Lee11a, LeePhD}
\begin{align}
\label{ewald}
	v(\theta_{ij}) & = \lim_{K \to \infty} v_K(\theta_{ij}),
	&
	v_K(\theta_{ij}) & = \frac{1}{\theta_{ij}} + \frac{1}{\pi}\sum_{k=1}^K \frac{k}{\left(k - \frac{\theta_{ij}}{2\pi} \right) \left(k +\frac{\theta_{ij}}{2\pi}\right)}.
\end{align}

The electron-electron interaction in Eq.~\eqref{ewald} diverges when $\theta_{ij} \to 0$. 
In higher dimensions, as shown by Kato \cite{Kato57}, the divergence of the interaction energy is cancelled by an equal and opposite divergence in the kinetic energy.
In 1D systems, the curvature of the wave function is unable to compensate for the divergence in the \alert{interaction potential, but the divergence in the energy is.}
This implies that the electronic wave function has nodes at all coalescence points (i.e. when two electrons touch). 
In other words, additionally to the Pauli exclusion principle which states that two electrons with same spin cannot occupy the same state, the singularity of the Coulomb operator enforces an opposite-spin version of the Pauli exclusion principle which also forbids two electrons with opposite spin to occupy the same quantum state.
Consequently, as previously mentioned \cite{Astrakharchik11, Lee11a, QR12, UEG1D, Ringium13}, the energy is independent of the spin-state for 1D systems and so we assume that all electrons are spin-up.

Note that, contrary to previous studies using quasi-1D Coulomb operator with a transverse harmonic potential \cite{Casula06, Shulenburger08, Shulenburger09, Lee11a}, here we use a strict-1D Coulomb interaction. (A comparison of the two types of interaction can be found in Ref.~\cite{Lee11a}.)
In the present study, we eschew the usual fictitious uniform positive background potential because its inclusion does not prevent a divergence of the Coulomb energy in 1D systems. \cite{Ringium13}. 
As shown below, the divergences induced by the infinite Ewald interaction ($K \to \infty$) and the thermodynamic limit ($n \to \infty$) can be rigorously handled.

%-----------------------------------------
\subsection{Hole curvature}
%-----------------------------------------
In the case of uniform electronic systems like the ones considered here, both $\rho$ and $\eta$ are constant, and we have $\tau_\text{IUEG}= (\pi^2/6) \rho^3$ in 1D. Thus, using Eqs.~\eqref{eta-def} and \eqref{orbitals}, one finds \cite{GLDA1} 
\begin{equation} 
\label{eta}
	\eta = 1 - 1/n^2.
\end{equation}
As one can see, there is a one-to-one mapping between the electron number $n$ and the two-electron $\eta$ parameter. In other words, $\eta$ gives information on the ``finiteness'' of the UEG.
Two important points have to be noted. First, in the thermodynamic limit ($\eta \to 1$), the present model is equivalent to the usual IUEG \cite{Lee11a}. 
Second, because we only consider the ground-state properties of finite (i.e. $n \ge 1$) and infinite (i.e. $n = \infty$) UEGs, the present model only covers the range $0 \le \eta \le 1$, as shown by Eq.~\eqref{eta}. 
Higher values of the hole curvature can be obtained by considering excited states. 
We are currently working on the extension of the GLDA based on UEG excited states \cite{GLDA14}.

%-----------------------------------------
\subsection{Hartree-Fock theory}
%-----------------------------------------
Due to the homogeneity of the system, it is straightforward to show that the HF wave function is a Slater determinant built on the occupied non-interacting orbitals \cite{Mitas06, Ringium13}
\begin{equation}
\label{PsiHF}
	\Psi_\text{HF}(\theta_1,\ldots,\theta_n) = \frac{1}{\sqrt{n!}} \left| \psi_{-(n-1)/2}(\theta_1) \ldots \psi_{(n-1)/2}(\theta_n) \right>,
\end{equation}
and the corresponding reduced HF energy is \cite{VignaleBook}
\begin{equation}
\label{EHF}
	\EHF(r_s,n) = \frac{\eps_{-2}(n)}{r_s^2} + \frac{\eps_{-1}(n)}{r_s},
\end{equation}
where $\eps_{-2}(n)$ represents the non-interacting kinetic energy and $\eps_{-1}(n)$ is the sum of the Coulomb and exchange energies:
\begin{align}
	\label{eps-2}
	\eps_{-2}(n) & = \frac{\pi^2}{n^3} \sum_{a}^\text{occ} \frac{a^2}{2} = \frac{n^2-1}{n^2} \frac{\pi^2}{24},
	\\
	\label{eps-1}
	\eps_{-1}(n) & = \frac{\pi}{n^2} \sum_{a < b}^\text{occ} \db{ab}{ab} = \frac{\pi}{n^2} \sum_{p=1}^{n-1} (n-p) V_p(K).
\end{align}
The double-bar integrals are 
\begin{equation}
\begin{split}
	\db{ab}{cd} 
	& = \iint \chi_a^*(\theta_1)\chi_b^*(\theta_2) v(\theta_{12}) \left[ \chi_c(\theta_1)\chi_d(\theta_2)-\chi_d(\theta_1)\chi_c(\theta_2)\right]\,d\theta_1\,d\theta_2
	\\
	& = 
	\begin{cases}
		V_{c-b}(K) - V_{c-a}(K),	&	\text{if $a+b=c+d$},
		\\
		0,				&	\text{otherwise},
	\end{cases}
\end{split}
\end{equation}
and $V_p(K)$ can be obtained in closed form for any value of the truncation order $K$:
 \begin{equation}
 \label{Vp}
	V_p(K) = \frac{\ln \left[p(2K+1)\,\pi\right] + \gamma - \Ci\left[p(2K+1)\,\pi\right]}{\pi},
\end{equation}
where $\gamma$ is the Euler-Mascheroni constant and $\Ci$ is the cosine integral \cite{NISTbook}. 
Note that, in 1D, both the Coulomb and exchange energies diverge logarithmically with opposite rate due to the singularity of the Coulomb operator for small interelectronic distances \cite{Ringium13}. 
Thus, they have to be consider together to ensure a finite result. For the infinite Ewald interaction ($K \to \infty$), Eq.~\eqref{Vp} simplifies to
\begin{equation}
	V_p = \lim_{K\to\infty}V_p(K) = \frac{\ln(2\pi p) + \gamma}{\pi},
\end{equation}
and yields, for large $K$, 
\begin{equation}
	\label{largeK-EHF}
	\eps_{-1}(n) \sim \frac{n-1}{n} \ln \sqrt{K} + \frac{\pi}{n^2} \sum_{p=1}^{n-1} (n-p) V_p + \ldots,
\end{equation} 
while, in the combined large-$n$ and large-$K$ limit, the HF potential energy behaves as
\begin{equation}
	\label{largeKn-EHF}
	\eps_{-1} \sim \ln \sqrt{K} + \ln \sqrt{n} + \frac{\gamma+\ln(2\pi)}{2} - \frac{3}{4} + \ldots.
\end{equation} 

%-----------------------------------------
\subsection{Correlation energy}
%-----------------------------------------

%%% TABLE 1 %%%
\begin{turnpage}
\begin{table*}
\caption{
\label{tab:Ec}
$-\Ec(\rs,\eta)$ (in $\mEh$ per electron) for the ground state of $n$ electrons on an infinitely thin wire with PBC. Statistical errors are shown in parentheses.} 
\
\begin{ruledtabular}
\begin{tabular}{ccd{1}d{1}d{1}d{1}d{1}d{1}d{1}d{1}d{1}d{1}d{1}d{1}}
			&			&														\mc{11}{c}{Wigner-Seitz radius $\rs = 1/(2\rho)$}														\\
																					\cline{3-13}
	\mc{1}{c}{$n$}		&	\mc{1}{c}{$\eta$}	&	\mc{1}{c}{0}		&	\mc{1}{c}{1/10}		&	\mc{1}{c}{1/5}		&	\mc{1}{c}{1/2}		&	\mc{1}{c}{1}		&	\mc{1}{c}{2}		&	\mc{1}{c}{5}		&	\mc{1}{c}{10}		&	\mc{1}{c}{20}		&	\mc{1}{c}{50}		&	\mc{1}{c}{100}		\\
	\hline
	2		&	3/4		&	14.168	&	13.914(0)	&	13.679(0)	&	13.011(0)	&	12.032(0)	&	10.463(0)	&   	7.563(0)	&	5.236(0)	&	3.303(0)	&	1.619(0)	&	0.894(0)	\\
	3		&	8/9		&	19.373	&	18.962(2)	&	18.581(0)	&	17.526(0)	&	16.031(0)	&	13.739(0)	&   	9.735(0)	&	6.662(0)	&	4.170(0)	&	2.030(0)	&	1.119(0)	\\
	4		&	15/16	&	21.917	&	21.404(5)	&	20.939(2)	&	19.657(1)	&	17.873(0)	&	15.205(0)	&	10.671(0)	&	7.265(0)	&	4.531(0)	&	2.199(0)	&	1.210(0)	\\
	5		&	24/25	&	23.373	&	22.804(3)	&	22.272(3)	&	20.845(1)	&	18.886(0)	&	15.997(0)	&	11.166(0)	&	7.579(0)	&	4.717(0)	&	2.286(0)	&	1.257(0)	\\
	6		&	35/36	&	24.293	&	23.672(3)	&	23.109(2)	&	21.582(1)	&	19.508(0)	&	16.477(0)	&	11.462(0)	&	7.765(0)	&	4.827(0)	&	2.336(0)	&	1.284(0)	\\
	7		&	48/49	&	24.916	&	24.270(2)	&	23.669(2)	&	22.075(1)	&	19.919(0)	&	16.792(0)	&	11.654(0)	&	7.885(0)	&	4.897(0)	&	2.369(0)	&	1.301(0)	\\
	8		&	63/64	&	25.361	&	24.686(3)	&	24.070(2)	&	22.421(1)	&	20.208(0)	&	17.011(0)	&	11.786(0)	&	7.967(0)	&	4.945(0)	&	2.391(0)	&	1.313(0)	\\
	9		&	80/81	&	25.691	&	24.996(3)	&	24.363(1) 	&	22.676(1)	&	20.418(0)	&	17.170(0)	&	11.881(0)	&	8.026(0)	&	4.979(0)	&	2.407(0)	&	1.321(0)	\\
	10		&	99/100	&	25.943	&	25.229(3)	&	24.588(2)	&	22.870(1)	&	20.577(0)	&	17.289(0)	&	11.952(0)	&	8.070(0)	&	5.005(0)	&	2.416(0)	&	1.328(0)	\\
	$\infty$	&	1		&	27.416	&	26.597	&	25.91(1)	&	23.962(1)	&	21.444(0)	&	17.922(0)	&	12.318(0)	&	8.292(0)	&	5.133(0)	&	2.476(0)	&	1.358(0)	\\
\end{tabular}	
\end{ruledtabular}
\end{table*}
\end{turnpage}

Our primary goal here is to determine the correlation functional $\Ec(r_s,\eta)$, or equivalently $\Ec(r_s,n)$, defined as
\begin{equation}
	\Ec(r_s,n)= \eps(r_s,n) - \EHF(r_s,n),
\end{equation}
where $\eps(r_s,n)$ is the exact reduced energy of the system and $\EHF(r_s,n)$ is defined in \eqref{EHF}. 
To build the GLDA correlation functional, we are going to combine information from the high-, intermediate- and low-density regimes.

%-----------------------------------------
\subsubsection{High density}
%-----------------------------------------
In the high-density regime ($r_s \ll 1$), the energy is expanded as a power series in terms of $r_s$ \cite{UEG1D}
\begin{equation}
	\eps(r_s,n) = \frac{\eps_{-2}(n)}{r_s^2} + \frac{\eps_{-1}(n)}{r_s} + \eps_0(n) + O(r_s).
\end{equation}
Thus, the high-density limiting correlation energy is
\begin{equation}
\label{Ec-HDL}
	\Ec(r_s,n) = \eps_0(n) + O(r_s),
\end{equation}
and
\begin{equation}
\label{eps_0}
	\eps_0(n) = - \frac{1}{n} \sum_{a<b}^{\text{occ}}\sum_{r<s}^{\text{virt}} \frac{ \db{ab}{rs} \db{rs}{ab}}{\kappa_r + \kappa_s - \kappa_a - \kappa_b}
\end{equation}
is given by second-order Rayleigh-Schr{\"o}dinger perturbation theory, and the summation over the virtual orbitals is 
\begin{equation}
	r = -\infty, \ldots, - \frac{n+3}{2}, - \frac{n+1}{2} \text{ and } +\frac{n+1}{2}, +\frac{n+3}{2}, \ldots, +\infty.
\end{equation}
$\eps_0(n)$ gives the \textit{exact} correlation energy at $r_s = 0$, which is a very valuable information. 
The values of $\eps_0(n)$ for various $n$ are reported in the $r_s = 0$ column of Table \ref{tab:Ec}.

In the combined high-density ($r_s \ll 1$) and thermodynamic ($n \to \infty$) limit, it is straightforward to show that \cite{UEG1D}
\begin{equation}
\label{HDL-thermo}
	\eps_0 = - \frac{1}{3\pi^2} \int_0^1 \int_{-x}^x \frac{\ln^3 \left(\frac{1+x}{1+y}\right) }{x-y} \,dy \,dx = -\frac{\pi^2}{360}.
\end{equation}

%-----------------------------------------
\subsubsection{Intermediate density}
%-----------------------------------------

For intermediate densities, we have performed diffusion Monte Carlo (DMC) \cite{Kalos74, Ceperley79, Reynolds82} calculations using the CASINO software \cite{CASINO10}. 
The results are reported in Table \ref{tab:Ec} for various $r_s$ and $\eta$ values.

The DMC energies for the IUEG are taken from Refs.~\cite{Lee11a}, \cite{UEG1D} and \cite{Ringium13}. 
The DMC calculations are performed with a population of approximately 1000 walkers and a time-step  $\tau = 0.008\,r_s^2$ following the Lee-Drummond methodology \cite{Lee11a}.
The trial wave function is of the Slater-Jastrow-backflow form \cite{Jastrow55}: the Slater determinant is the HF wave function and we use a backflow transformation to evaluate the orbitals \cite{Kwon93, LopezRios06, Lee11b}. 
The Jastrow factor includes two-body terms while the backflow transformation provides an efficient way of describing three-body effects \cite{Holzmann03}.  
For $r_s \ge 50$, only variational Monte Carlo (VMC) \cite{McMillan65, Ceperley77, Umrigar99} are required to reach microhartree accuracy.
Note that, because the nodes of the HF wave function are exact for the ground state and the backflow transformation leaves the nodes unchanged in the present case, there is no fixed-node errors in our QMC calculations \cite{Astrakharchik11, Lee11a, Ringium13}. 
Thus, DMC energies are actually exact within statistical errors.

%-----------------------------------------
\subsubsection{Low density}
%-----------------------------------------

In the low-density regime ($r_s \gg 1$), the system crystallize to form a so-called Wigner crystal (WC) \cite{Wigner34}. 
In 1D, the WC consists of $n$ electrons separated by a distance  $L/n$ \cite{Fogler05a} or equivalently an angle $2\pi/n$ \cite{Ringium13}. 
Using strong-coupling perturbation theory, the energy is expanded in terms of $r_s$ and reads 
\begin{equation}
	\eps(r_s,n)  = \frac{\eps_\text{WC}(n)}{r_s} + O(r_s^{-3/2})	,
\end{equation}
where
\begin{equation}
\label{EWC}
	\eps_\text{WC}(n) = \frac{\pi}{n^2} \sum_{i<j}^{n} v(\Hat{\theta}_{ij})
\end{equation} 
is the classical energy of the WC, and $\Hat{\theta}_{ij} = \left| i - j \right| 2\pi/n$ are the equilibrium interelectronic angles between electrons $i$ and $j$ in the crystal. 
Equation \eqref{EWC} simplifies as
\begin{equation}
	\eps_\text{WC}(n) = \frac{H_{n-1}}{2} + \frac{1}{2n^2} \sum_{p=1}^{n-1} (n-p) 
	\left( H_{K+p/n} + H_{K-p/n} - H_{p/n} - H_{-p/n} \right),
\end{equation}
where $H_n$ is a harmonic number \cite{NISTbook}. For large $K$, we find
\begin{equation}
\label{largeK-eW}
	\eps_\text{WC} \sim \frac{n-1}{n} \ln \sqrt{K} + \frac{H_n -1}{2} 
	- \frac{1}{2n^2} \sum_{p=1}^{n-1} (n-p) \left( \frac{H_{p/n} + H_{-p/n}}{2} - \gamma \right)+ \ldots,
\end{equation}
while, for large $n$ and large $K$, we have
\begin{equation}
\label{largeKn-eW}
	\eps_\text{WC}(n) \sim \ln \sqrt{n} + \ln \sqrt{K} + \frac{\gamma}{2}.
\end{equation}
Equations \eqref{largeK-eW} and \eqref{largeKn-eW} exhibit the same logarithmic divergences as the HF energy in Eqs.~\eqref{largeK-EHF} and \eqref{largeKn-EHF}. Thus, the low-density correlation energy expansion
\begin{equation}
\label{Ec-LDL}
	\Ec(r_s,n) = \frac{\eps_\text{W}(n)-\eps_{-1}(n)}{r_s} + O(r_s^{-3/2}) = \frac{\eps_\infty(n)}{r_s} + O(r_s^{-3/2})
\end{equation}
is finite for any number of electrons:
\begin{equation}
\label{eps_infty}
	\eps_\infty(n) =  \frac{H_n -1}{2}  
	- \frac{1}{2n^2} \sum_{p=1}^{n-1} (n-p) \left[ \frac{H_{p/n} + H_{-p/n}}{2} + \ln(2\pi p)  \right].
\end{equation}
As first shown by Fogler \cite{Fogler05a}, in the combined low-density ($r_s \gg 1$) and thermodynamic ($n \to \infty$) limit, the correlation energy is 
\begin{equation}
\label{LDL-thermo}
	\Ec(r_s) = \frac{\ln{\sqrt{2\pi}} - 3/4}{r_s} + O(r_s^{-3/2}).
\end{equation}

%--------------------------------
\section{
\label{sec:GLDA}
gLDA functional}
%---------------------------------

%%% TABLE 2 %%%
\begin{table}
	\caption{
	\label{tab:coeffs}
	Coefficients of $\Upsilon_0(\eta)$ $\Upsilon(\eta)$ and $\Upsilon_\infty(\eta)$ for the gLDAw and rev-gLDAr functionals.} 
	\begin{ruledtabular}
		\begin{tabular}{cd{-1}d{-1}d{-1}cd{-1}d{-1}d{-1}}
						&	\mc{3}{c}{gLDAw}																		&&	\mc{3}{c}{rev-gLDAr}
						\\
						\cline{2-4} 																				\cline{6-8}
						&	\mc{1}{c}{$\Upsilon_0(\eta)$}	&	\mc{1}{c}{$\Upsilon(\eta)$}	&	\mc{1}{c}{$\Upsilon_\infty(\eta)$}	&&	\mc{1}{c}{$\Upsilon_0(\eta)$}	&	\mc{1}{c}{$\Upsilon(\eta)$}	&	\mc{1}{c}{$\Upsilon_\infty(\eta)$}	\\
			\hline
			$c_1$		&	0.025979					&	33.0265					&		0.163723					&&	0.025873					&	18.3407					&	0.164037		\\
			$c_2$		&	0.025979					&	0.896251					&		0.163723					&&	0.025873					&	-0.154372					&	0.164037		\\
			$c_3$		&	0.033891					&	24.2518					&		0.301135					&&	0.032541					&	13.2193					&	0.261152		\\
			$c_4$		&	0.642367					&	16.1820					&		0.661217					&&	0.741760					&	8.807757					&	0.519097		\\
			$c_5$		&	-0.35379					&	-12.5392					&		0.152167					&&	-0.498560					&	-6.681718					&	0.055756		\\
		\end{tabular}
	\end{ruledtabular}
\end{table}

We have now all the information required to construct the GLDA correlation functional. Because the present approach does only cover the range $0 \le \eta \le 1$ (see above) and is based on the infinitely-thin-wire model, we name the present functional $\Ec^\text{gLDAw}$ and we define it as follow
\begin{equation} 
\label{kergLDA}
	\Ec^\text{gLDAw}(\rs,\eta) =	
		\begin{cases}
			\Ec^\text{GLDAw}(\rs,\eta),					&	\eta < 1	\\
			\Ec^\text{LDA}(\rs) = \Ec^\text{GLDAw}(\rs,1),	&	\eta \ge 1
		\end{cases}
\end{equation}
with
\begin{equation}
\label{GLDA}
	\Ec^\text{GLDAw}(\rs,\eta) = \Upsilon_0(\eta)\,F\left[1, \frac{3}{2}, \Upsilon(\eta) , \frac{2\Upsilon_0(\eta)(1-\Upsilon(\eta))}{\Upsilon_\infty(\eta)}\,\rs \right].
\end{equation}
In \eqref{GLDA}, $F(a, b, c,x)$ is the Gauss hypergeometric function \cite{NISTbook} chosen to make sure that $\Ec^\text{GLDAw}(\rs,\eta)$ exactly reproduces the behavior of the correlation energy at high (Eq.~\eqref{Ec-HDL}) and low (Eq.~\eqref{Ec-LDL}) densities. 
The functions $\Upsilon_0(\eta)$ and $\Upsilon_\infty(\eta)$ are obtained by fitting the high-density $\eps_0(n)$ (Eq.~\eqref{eps_0} and Table \ref{tab:Ec}) and low-density $\eps_\infty(n)$ (Eq.~\eqref{eps_infty}) functions, respectively, while $\Upsilon(\eta)$ is determined using the intermediate-density correlation energies gathered in Table \ref{tab:Ec}. 
They are all approximated using the same functional form
\begin{equation}
\label{Upsilon-functions}
	\Upsilon_0(\eta),\,\Upsilon(\eta),\,\Upsilon_\infty(\eta) = \frac{c_1 - c_2 \sqrt{1-\eta} - c_3\,\eta}{c_4 + \sqrt{1-\eta} + c_5\,\eta},
\end{equation}
where the limiting behavior of $\Upsilon_0(\eta)$ and $\Upsilon_\infty(\eta)$ are fixed to their exact values i.e. $\Upsilon_0(0) = \Upsilon_\infty(0) = 0$, $\Upsilon_0(1) = -\pi^2/360$ (Eq.~\eqref{HDL-thermo}) and $\Upsilon_\infty(1)=\ln{\sqrt{2\pi} - 3/4}$ (Eq.~\eqref{LDL-thermo}). 
In terms of $n$, Eq.~\eqref{Upsilon-functions} is a Pad\'e approximant which has been determined to reproduce the behavior of the high- and low-density correlation energies at small and large $n$ \footnote{In a previous study \cite{GLDA1}, the functions $\Upsilon_0(\eta)$, $\Upsilon(\eta)$ and $\Upsilon_\infty(\eta)$ were obtained using truncated series or even fit of truncated series, and they have shown to provide slightly less robust results.}.

The coefficients $c_i$ can be found in Table \ref{tab:coeffs} for each function. Although we have used a very limited amount of information from the high- intermediate- and low-density regimes to construct $\Ec^\text{GLDAw}(\rs,\eta)$, the gLDA functional \eqref{GLDA} is extremely robust, with maximum and mean errors of $0.1$ and $0.03$ $\mEh$, respectively, compared to the DMC values gathered in Table \ref{tab:Ec}. Note that, by construction, the correlation energy of \textit{any} one-electron system is zero.

%--------------------------
\section{
\label{sec:res}
Results and discussion}
%--------------------------

%%% TABLE 3 %%%
\begin{table*}
	\caption{
	\label{tab:apps}
	$-E_\text{c}$ (in $\mEh$) of $n$-boxium and $n$-hookium for $L=\pi$ and $k=1$ with $n=$ 2, 3, 4, 5, and 6.} 
	\begin{ruledtabular}
		\begin{tabular}{ld{1}d{1}d{1}d{1}d{1}d{1}d{1}d{1}d{1}d{1}}
						&	\mc{5}{c}{$n$-boxium ($L=\pi$)}								&	\mc{5}{c}{$n$-hookium ($k = 1$)}								\\
						\cline{2-6}														\cline{7-11}
						&	\mc{1}{l}{$n = 2$}	&	\mc{1}{c}{$n = 3$}	&	\mc{1}{c}{$n = 4$}	&	\mc{1}{c}{$n = 5$}	&	\mc{1}{c}{$n = 6$}	
						&	\mc{1}{c}{$n = 2$}	&	\mc{1}{c}{$n = 3$}	&	\mc{1}{c}{$n = 4$}	&	\mc{1}{c}{$n = 5$}	&	\mc{1}{c}{$n = 6$}						\\
			\hline
			LDA			&	46.0		&	72.5		&	99.3		&	126.4	&	154	&	42.1		&	65.8		&	90.0		&	114.5	&	139		\\
			gLDAr		&	10.9		&	26.3		&	43.9		&	63.0		&	83	&	12.7		&	27.9		&	44.8		&	62.8		&	82		\\
			rev-gLDAr		&	11.0		&	26.5		&	44.2		&	63.3		&	83	&	12.8		&	28.1		&	45.0		&	63.1		&	82		\\
			gLDAw		&	11.3		&	27.1		&	45.3		&	64.9		&	86	&	13.1		&	28.9		&	46.3		&	64.8		&	84		\\
			FCI 			&	9.8		&	26.2		&	46.1		&	68.0		&	92	&	13.5		&	31.8		&	52.4		&	74.3		&	101		\\
		\end{tabular}
	\end{ruledtabular}
\end{table*}

%%% TABLE 4 %%%
\begin{table*}
	\caption{
	\label{tab:apps2}
	$-E_\text{c}$ (in $\mEh$) of 2-boxium (Bo) and 2-hookium (Ho) as a function of $L$ or $k$.} 
	\begin{ruledtabular}
		\begin{tabular}{ccccccccccc}
			$L/\pi$ or $k^{-1/4}$	&	\mc{2}{c}{LDA}			&	\mc{2}{c}{gLDAr}	&	\mc{2}{c}{rev-gLDAr}	&	\mc{2}{c}{gLDAw}		&	\mc{2}{c}{Exact}		\\
			\hline
							&	Bo		&	Ho		&	Bo		&	Ho		&	Bo		&	Ho			&	Bo		&	Ho		&	Bo		&	Ho		\\
							\cline{2-3}					\cline{4-5}					\cline{6-7}						\cline{8-9}					\cline{10-11}
			$1/8$			&	53.4		&	52.5		&	15.2		&	15.2		&	15.3		&	15.3			&	15.7		&	19.7		&	13.7		&	18.8		\\
			$1/4$			&	52.2		&	50.6		&	14.5		&	17.8		&	14.6		&	18.0			&	15.0		&	18.6		&	13.1		&	18.0		\\		
			$1/2$			&	49.9		&	47.4		&	13.2		&	15.9		&	13.3		&	16.0			&	13.7		&	16.6		&	11.9		&	16.4		\\		
			$1$				&	46.0		&	42.1		&	10.9		&	10.9		&	11.0		&	11.0			&	11.3		&	13.1		&	9.8		&	13.6		\\			
			$2$				&	40.1		&	34.6		&	7.4		&	7.9		&	7.4		&	7.9			&	7.6		&	8.1		&	6.7		&	9.1		\\
			$4$				&	32.8		&	25.9		&	3.5		&	3.0		&	3.5		&	2.9			&	3.6		&	3.0		&	3.3		&	4.2		\\
			$8$				&	25.2		&	17.9		&	1.0		&	0.6		&	1.0		&	0.6			&	1.0		&	0.6		&	1.0		&	1.2		\\
		\end{tabular}
	\end{ruledtabular}
\end{table*}

%%% FIGURE 1 %%%
\begin{figure}
	\includegraphics[height=0.35\textheight]{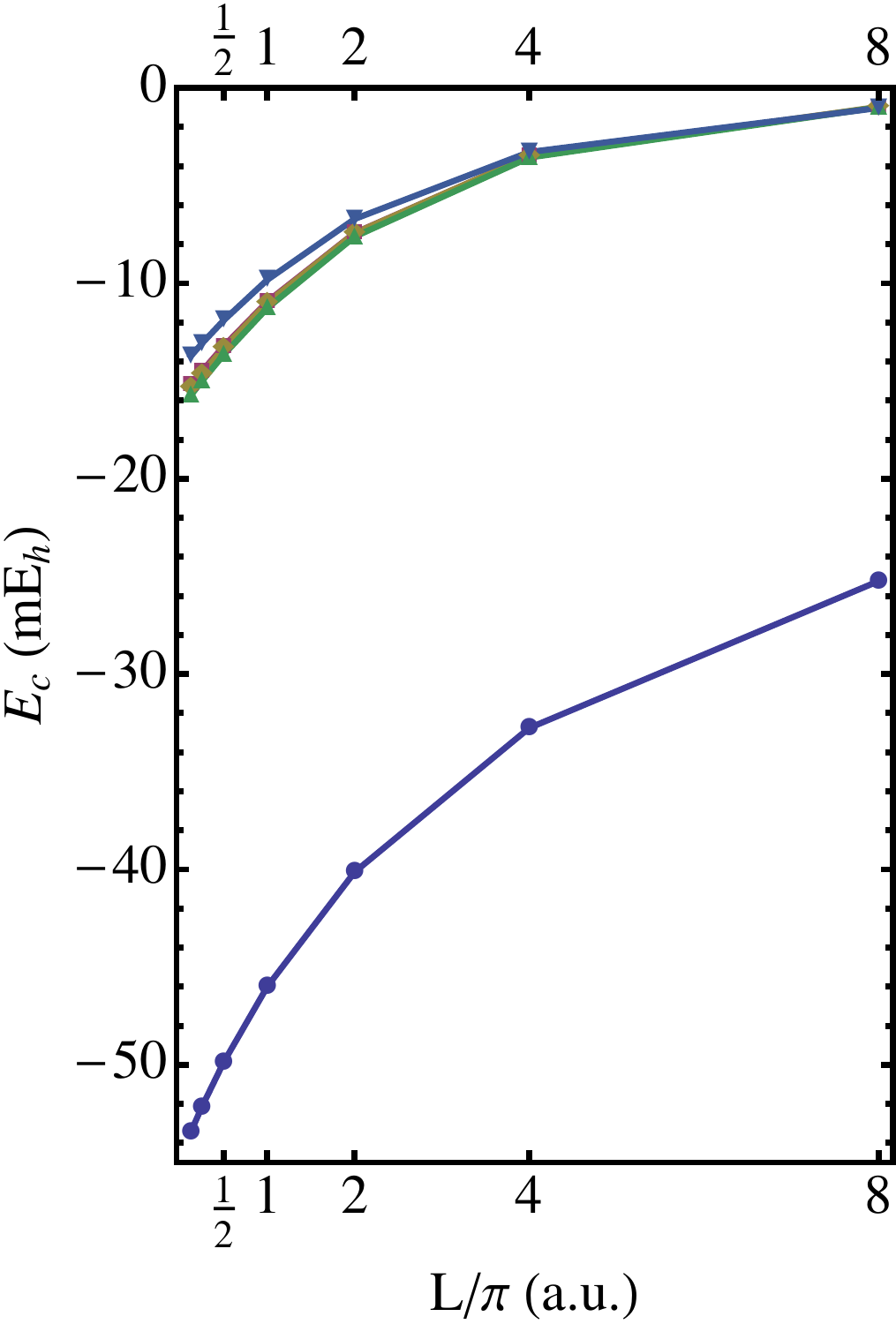}
	\includegraphics[height=0.35\textheight]{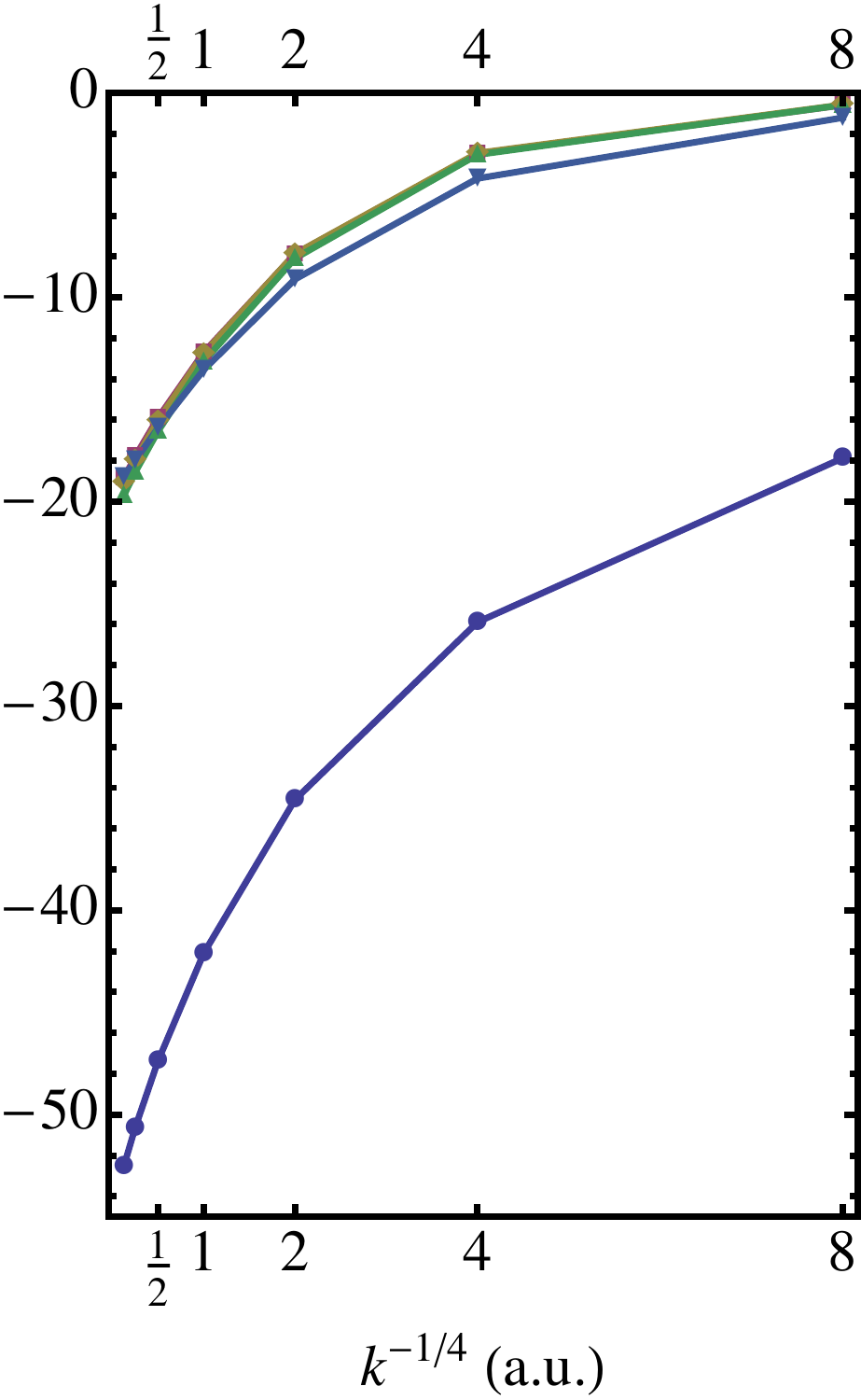}
	\includegraphics[height=0.35\textheight]{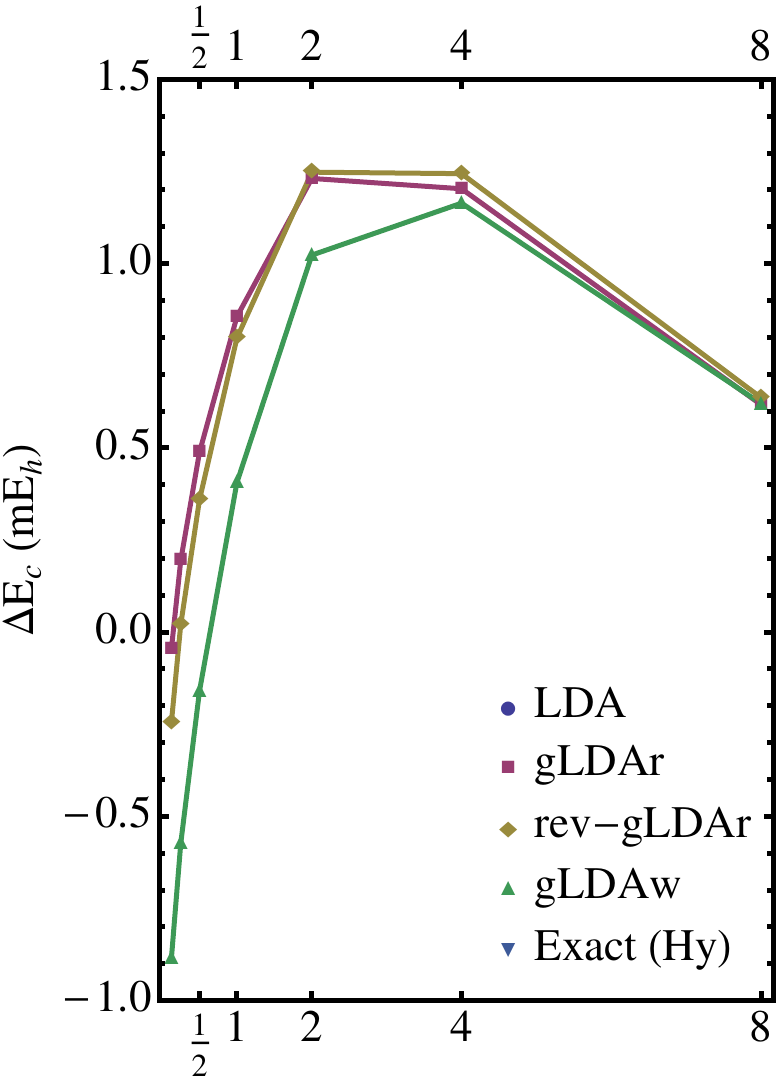}
	\includegraphics[height=0.35\textheight]{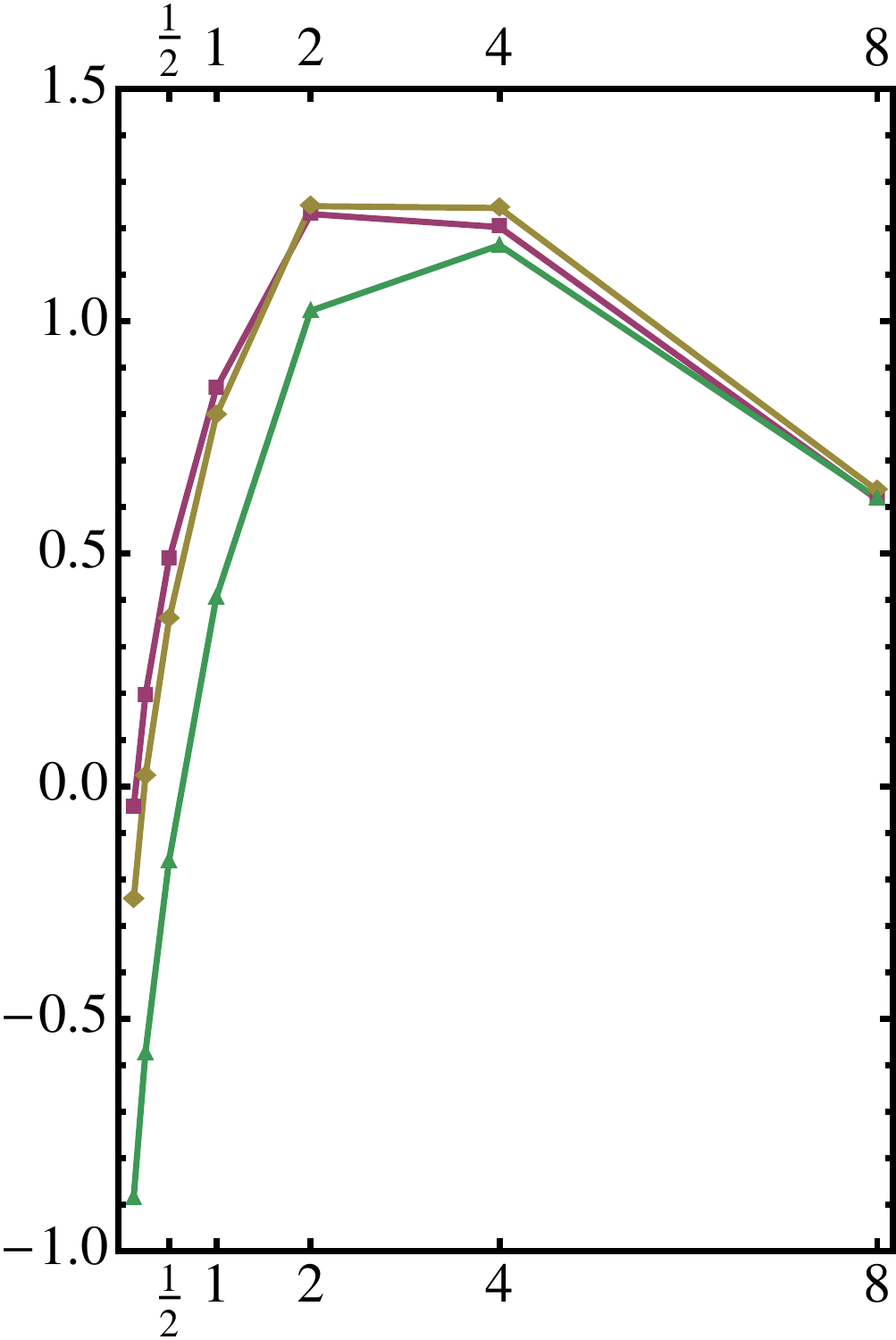}
	\caption{
	\label{fig:bo-ho}
	\alert{(Color online)} Top left: $E_\text{c}$ (in $\mEh$) of 2-boxium as a function of $L$. 
	Top right: $E_\text{c}$ (in $\mEh$) of 2-hookium as a function of $k$. 
	Bottom left: $\Delta E_\text{c} = E_\text{c} - E_\text{c}^\text{Hy}$ (in $\mEh$) of 2-boxium as a function of $L$. 
	Bottom right: $\Delta E_\text{c} = E_\text{c} - E_\text{c}^\text{Hy}$ (in $\mEh$) of 2-hookium as a function of $k$.}
\end{figure}

To demonstrate the performance of the gLDA functional defined in \eqref{kergLDA}, we compute the ground state correlation energy of various inhomogeneous systems. 
The two systems considered here consist of $n$ spin-up electrons in a box of length $L$, and a harmonic well of force constant $k$. 
We call these systems $n$-boxium and $n$-hookium, respectively (see Ref.~\cite{GLDA1} for more details). 
The LDA, gLDAr and gLDAw calculations corresponds to non-self-consistent post-HF calculations \footnotemark[1] based on the LDA functional (see Eq.~\eqref{kergLDA})
\begin{equation}
	\Ec^\text{LDA}(\rs) = \Ec^\text{GLDAw}(\rs,1),
\end{equation}
the ``ring-based'' gLDA functional constructed by Loos, Ball and Gill in Ref.~\cite{GLDA1}, and the present ``wire-based'' gLDA functional defined in Eq.~\eqref{kergLDA}, respectively. 
We have also re-parametrized the gLDAr functional using strictly the same approach as in Sec.~\ref{sec:GLDA} but based on the data of Ref.~\cite{GLDA1}. 
We call this new functional rev-gLDAr and we report the coefficients of $\Upsilon_0(\eta)$, $\Upsilon(\eta)$ and $\Upsilon_\infty(\eta)$ in Table \ref{tab:coeffs}. 
The new rev-gLDAr fit is shown to be more robust than the previous gLDAr fit with a maximum error of 0.1 $\mEh$ and a mean error of $0.03$ $\mEh$ compared to the benchmark energies of Ref.~\cite{GLDA1}.

In Table \ref{tab:apps}, we have reported the total correlation energy $E_\text{c}$ of $n$-boxium and $n$-hookium for $L=\pi$ and $k=1$ with $n=$ 2, 3, 4, 5, and 6. 
The exact energies are obtained with full configuration interaction (FCI) calculations \cite{Knowles84, Knowles89}. 
The results show that the three gLDA functionals perform exceptionally well compared to the exact FCI values. 
Except for 2- and 3-boxium, the present gLDAw functional yields more accurate results than gLDAr and rev-gLDAr.
Note that the re-parametrized rev-gLDAr functional only slightly improve (by ca. 0.2 or 0.3 $\mEh$) the correlation energy values.
The LDA functional strongly overestimates the correlation energies for small number of electrons. 
This trend is attenuated for larger $n$ as expected (the LDA becomes exact in the large-$n$ limit).
For these two systems, the ring-based model is probably less appropriate than the wire-based model due to the interaction of the electrons ``through'' the ring.

In Table \ref{tab:apps2}, we compute the correlation energy $E_\text{c}$ of 2-boxium and 2-hookium for various values of the box length $L$ and the harmonic force constant $k$, respectively. 
For large $L$ or small $k$, the Coulomb interaction is dominant and these systems are strongly correlated. In the opposite regime (i.e. small $L$ or large $k$), the kinetic energy dominates and these systems are said to be weakly correlated.
Our benchmark values are based on near-exact Hylleraas (Hy) calculations \cite{Hylleraas29, Hylleraas30, Hylleraas64}. 
These results are depicted in Fig.~\ref{fig:bo-ho}. 
In the two bottom graphs of Fig.~\ref{fig:bo-ho}, we have plotted the difference in correlation energy $\Delta E_\text{c} = E_\text{c} - E_\text{c}^\text{Hy}$ for the three gLDA functionals.

The results show that the gLDA functionals yield accurate results for the weak, intermediate and strong regimes of correlation with a maximum error of 1.3 $\mEh$, while the LDA yields very poor estimates of the correlation energy for each regime.
Even for strongly correlated systems where DFT functionals usually fail \cite{Cohen12} (see Refs.~\cite{GoriGiorgi09, Francesc12} for alternative approaches), the gLDA functionals behave exceptionally well compared to the near-exact Hylleraas results. 
Again, except at very high density, the gLDAw functional gives more accurate results than the ringium-based gLDA functionals (gLDAr and rev-gLDAr). 
However, the difference between these values are rather small, which shows the weak system-dependence of the GLDA method.

%==============
\section{
\label{sec:ccl}
Conclusion}
%===============

In the present study, we have constructed a new generalized local-density approximation (GLDA) correlation functional based on finite uniform electron gases (UEGs) built by considering electrons on an infinitely thin wire with periodic boundary conditions. These UEGs are finite versions of the ubiquitous infinite homogeneous electron gas, the cornerstone of the most popular density functional paradigm \textemdash~the local-density approximation (LDA) \textemdash. 
We have performed a comprehensive study of these finite UEGs at high, intermediate and low densities using perturbation theory and quantum Monte Carlo calculations. 
We have shown that this new functional yields very robust correlation energies for various inhomogeneous one-dimensional systems in both the strongly- and weakly-correlated regimes. 

The present approach can be easily extended to higher dimensions by computing the exchange and correlation energies of finite uniform electron gases for various spin-polarization \cite{Perdew81,Ma11} using current QMC softwares \cite{CASINO10}. 
However, unlike the present case, the magnitude of the fixed-node errors has to be addressed and it has been shown that large differences in the fixed-node errors can appear  depending on the degree of node nonlinearity \cite{Rasch14}. In the case of large fixed-node errors, the performance of the FCI-QMC method developed by Alavi and coworkers \cite{Booth09, Booth10, Cleland10, Cleland11} will be investigated. This method has no fixed-node error and it has been successfully applied to 3D UEGs recently \cite{Shepherd12a, Shepherd12b, Shepherd12c}.

%======================
\begin{acknowledgments}
%======================
The author thanks the Australian Research Council for funding (Grant DE130101441 and DP140104071), the NCI National Facility for a generous grant of supercomputer time, and Amy Kendrick, Neil Drummond, Mike Towler and Peter Gill for stimulating discussions.
\end{acknowledgments}

\appendix

\section{
\label{app:ringium}
The ringium model}

The Hamiltonian of the system consisting of $n$ electrons on a ring of radius $R$ (that we have also called ringium) is \cite{Ringium13, GLDA1} 
\begin{equation}
\label{H-ringium}
	\Hat{H} = - \frac{1}{2R^2} \sum_{i=1}^{n} \frac{\partial^2}{\partial\theta_i^2} + \sum_{i<j}^{n} \frac{1}{r_{ij}},
\end{equation}
where $\theta_i$ is the angle of electron $i$ around the ring center,
\begin{equation}
\label{rij-ringium}
	r_{ij} = \left| \bm{r}_i - \bm{r}_j \right| = R \sqrt{2-2\cos \theta_{ij}}
\end{equation}
is the across-the-ring distance between electrons $i$ and $j$, and the interelectronic distance is $\theta_{ij} = \left| \theta_i - \theta_j \right|$.
Compared to the Hamitonian of $n$ electrons on an infinitely thin wire given by Eq.~\eqref{H-wire}, one can see that the only difference between the two systems is the electron-electron interaction. 
In ringium, the interelectronic potential $r_{ij}^{-1}$ is periodic and the electrons interact ``through the ring'' (see Eq.~\eqref{rij-ringium}). 
In the electrons-on-a-wire paradigm, we use the Ewald interaction potential given by \eqref{ewald} which corresponds to an infinite summation of the Coulomb interaction.
However, because the system interacts with its own images, finite-size errors arise.

It is interesting to compare the correlation energies obtained for various $n$ and $r_s$ values. 
Correlation energies for the wire-based and ring-based systems are gathered in Table \ref{tab:Ec} of the present manuscript and Table II of Ref.~\cite{GLDA1}, respectively. 
As one can see, for a given value of $n$ and $r_s$, the difference in correlation energy is very small. 
The largest difference is less than a millihartree for $n=2$ and $r_s = 0$.

\end{document}